\documentclass[
    pre,
    aps,
    amsmath,
    amssymb,
    reprint,
    superscriptaddress,
    nofootinbib
]{revtex4-2}

\usepackage{graphicx}
\usepackage{dcolumn}
\usepackage{bm}
\usepackage[
    bookmarks=false,
    colorlinks=true,
    linkcolor=blue,
    citecolor=blue,
    urlcolor=blue,
]{hyperref}
\usepackage{orcidlink}
\usepackage{titlesec}

\DeclareMathOperator{\sinhc}{sinhc}
\newcommand{\sperp}{\mathrel{\raisebox{0.1ex}{\scalebox{0.5}{$\perp$}}}}
\newcommand{\sparallel}{\mathrel{\raisebox{0.1ex}{\scalebox{0.5}{$\parallel$}}}}

\definecolor{aesthetic-background}{RGB}{30, 20, 40}
\definecolor{aesthetic-blue}{RGB}{0, 174, 255}
\definecolor{aesthetic-cyan}{RGB}{90, 200, 230}
\definecolor{aesthetic-green}{RGB}{55, 196, 55}
\definecolor{aesthetic-magenta}{RGB}{249, 42, 173}
\definecolor{aesthetic-yellow}{RGB}{253, 163, 42}
\definecolor{ppt-blue}{RGB}{2, 83, 118}

\makeatletter
\newcommand*\@secondofsix[6]{#2}
\newcommand{\addtotitleformat}{%
  \@ifstar{\addtotitleformat@star}{\addtotitleformat@nostar}}
\newcommand\addtotitleformat@nostar[2]{%
  \PackageError{titlesec}{non starred form of \string\addtotitleformat\space not supported}{}}
\newcommand\addtotitleformat@star[2]{%
  \expandafter\expandafter\expandafter\expandafter
  \expandafter\expandafter\expandafter\def
  \expandafter\expandafter\expandafter\expandafter
  \expandafter\expandafter\expandafter\@currentsection@font
  \expandafter\expandafter\expandafter\expandafter
  \expandafter\expandafter\expandafter{%
    \expandafter\expandafter\expandafter\@secondofsix
       \csname ttlf@\expandafter\@gobble\string#1\endcsname}%
  \titleformat*{#1}{\@currentsection@font#2}%
}
\makeatother

\addtotitleformat*{\section}{\MakeUppercase}

\makeatletter
\renewcommand*\date[1][\Dated@name]{
    \def\@date{
        #1\today
    }
}
\makeatother

\begin{document}


\title{
%
    Thermodynamic fluctuations in freely jointed chains under force
}
\author{
    Michael R. Buche%
    \:\orcidlink{0000-0003-1892-0502}\,
}
\email{mrbuche@sandia.gov}
\affiliation{
    Materials and Failure Modeling, Sandia National Laboratories, Albuquerque, New Mexico 87185, USA
}
\author{
    Alvin Chen%
    \:\orcidlink{0000-0001-5252-0372}\,
}
\affiliation{
    Materials and Failure Modeling, Sandia National Laboratories, Albuquerque, New Mexico 87185, USA
}
\date{}

\begin{abstract}
It is common to study polymer physics through the use of idealized single-chain models, and the most popular of these is the freely jointed chain model.
In certain thermodynamic ensembles, statistical mechanical treatment of this model is analytically tractable or sometimes exactly solvable.
This enables useful relations to be ascertained, like the expected chain end-to-end length as a function of an applied force.
However, most of these relations return ensemble averages, which are values with inherent uncertainty, as opposed to deterministic values with no variance.
This is an important distinction to understand and quantify, because the majority of studies to date involving single-chain models effectively treat these values as deterministic rather than fluctuating.
To address this issue, thermodynamic fluctuations are examined in the freely jointed chain model.
Specifically, the probability densities and standard deviations of the longitudinal, lateral, transverse, and radial portions of the chain extension, as well as the extension and link angles, are examined for different numbers of links and applied forces.
Fluctuations in these quantities are shown to be considerable until the applied force becomes large.
Increasing the number of links in the chain gradually reduces fluctuations in all quantities except for the link angles, since they are independent for freely jointed chains in the isotensional ensemble.
Quantities are obtained analytically whenever possible and numerically otherwise.
Overall, these results provide intuitive admonitions to consider when modeling the stretching of single polymer chains or the deformation of entire polymer networks.
\smallskip\smallskip\smallskip

\phantom{\noindent DOI: \href{https://doi.org/??.????/???????????.??????}{??.????/????????.???.??????}}
\end{abstract}

\maketitle


\section{Introduction}\label{sec:introduction}

The application of mechanical force is often used to probe the behavior of single polymer chains.
The chain resists extension according to how the applied force compares to the thermal energy $kT=1/\beta$, the strength of intramolecular interactions, and relevant length scales.
Before forces are sufficiently large enough to distort intramolecular potentials, this resistance arises primarily by the reduction of configurational entropy as the chain is extended.
The canonical single-chain model for describing this phenomenon is the freely jointed chain model, a set of rigid links connected by penalty-free hinges \cite{treloar1949physics}.
Under a constant applied force, which is known as the isotensional thermodynamic ensemble \cite{buche2022freely}, the expected extension of the chain conjugate to the applied force is exactly given by the Langevin function \cite{kuhn1942beziehungen}.
In any case, the measured value conjugate to the applied boundary condition is not truly a deterministic result.
Rather, the measurement is merely an ensemble average of a value that is continually fluctuating \cite{mcquarrie}.
Additionally, the extension calculated by the Langevin function for a freely jointed chain under force is, in fact, only the longitudinal component of the extension.
It is crucial to understand these nuances when modeling single-molecule stretching experiments \cite{wang1997stretching,rief1997single,janshoff2000force,frey2012understanding} and formulating physically-based constitutive models \cite{lavoie2019modeling,buche2021chain,mulderrig2021affine,lamont2021rate,grasinger2023polymer,mulderrig2025polydisperse}, which do motivate this study.
Fluctuations, which are rarely considered \cite{fiasconaro2019analytical}, and probability distributions are examined for the various components of a freely jointed chain extension under force.
Analytic relations are presented when available, where ten billion Monte Carlo samples spread amongst three hundred equally-spaced bins are used otherwise, calculated in either case using \texttt{conspire} \cite{conspire}.

\section{Chain extension}\label{sec:extension}

Central to the isotensional statistical thermodynamics of the freely jointed chain model is the partition function
\begin{equation}\label{eq:z}
Z(\eta) = \int d\Omega_1\cdots\int d\Omega_{N_b}\exp\left(\eta\sum_{i=1}^{N_b}\cos\theta_i\right),
\end{equation}
where $N_b$ is the number of links, $\ell_b$ is the link length, $\eta=\beta f\ell_b$ is the nondimensional force, $\Omega=\sin\theta\,d\theta\,d\phi$ is the differential solid angle, $\theta\in[0,\pi]$ is the polar angle, and $\phi\in[0,2\pi]$ is the azimuthal angle.
Note that the direction of the applied force has been chosen to be the positive polar axis.
Also note that the partition function in Eq.~\eqref{eq:z} can be written as $Z=z^{N_b}$, where $z\propto\sinh\eta/\eta$ is the partition function for one freely jointed link \cite{buche2022freely}.
The partition function can be utilized to calculate the isotensional ensemble average of some quantity $f$ as
\begin{equation}\label{eq:f}
\langle f\rangle = \frac{1}{Z(\eta)}\int d\Omega_1\cdots\int d\Omega_{N_b}\,f\exp\left(\eta\sum_{i=1}^{N_b}\cos\theta_i\right).
\end{equation}
In certain fortunate cases, these ensemble averages or other moments are related to derivatives of $\ln Z(\eta)$.

\subsection{Longitudinal extension}\label{sec:extension:longitudinal}

\begin{figure}[t]
\includegraphics{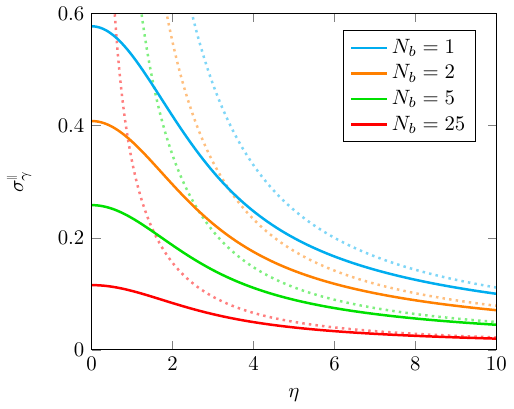}
\caption{\label{fig:gamma:longitudinal:s}%
Standard deviation $\sigma_\gamma^{\sparallel}$ of longitudinal extension $\gamma_{\sparallel}$ as a function of the applied nondimensional force $\eta$ for freely jointed chains with numbers of links $N_b$.
Dotted lines show $\sigma_\gamma^{\sparallel}$ relative to $\langle\gamma_{\sparallel}\rangle$, the ensemble average longitudinal extension.
}
\end{figure}

The nondimensional extension is defined as the chain end-to-end length scaled by the contour length $N_b\ell_b$, and here will be referred to as simply the extension.
The extension is generally formulated as a vector $\boldsymbol{\gamma}$ in some coordinate system relative to the applied force vector $\boldsymbol{\eta}$.
For instance, the longitudinal extension $\gamma_{\sparallel}$ is given by
\begin{equation}\label{eq:gamma:longitudinal}
\gamma_{\sparallel} = \frac{1}{N_b}\sum_{i=1}^{N_b}\cos\theta_i,
\end{equation}
which is the magnitude of $\boldsymbol{\gamma}$ that is coaxial with force $\boldsymbol{\eta}$.
The ensemble average of $\gamma_{\sparallel}$ via Eq.~\eqref{eq:f} is exactly given by a derivative of $\ln Z(\eta)$ as
\begin{equation}\label{eq:gamma:longitudinal:m}
\langle\gamma_{\sparallel}\rangle = \frac{1}{N_b}\frac{\partial\ln Z}{\partial\eta} = \mathcal{L}(\eta),
\end{equation}
where $\mathcal{L}(\eta)=\coth\eta-1/\eta$ is the Langevin function \cite{treloar1949physics}.
Fortunately, through calculating another derivative,
\begin{equation}
\frac{\partial\langle\gamma_{\sparallel}\rangle}{\partial\eta} = N_b\langle\gamma_{\sparallel}^2\rangle - N_b\langle\gamma_{\sparallel}\rangle^2,
\end{equation}
the variance in the longitudinal extension is $\mathcal{L}'(\eta)/N_b$, which means the standard deviation is
\begin{equation}\label{eq:gamma:longitudinal:s}
\sigma_\gamma^{\sparallel}(\eta) = \sqrt{\frac{1}{N_b}\left(\frac{1}{\eta^2} - \frac{1}{\sinh^2\!\eta}\right)}.
\end{equation}
Analytic relations for $\langle\gamma_{\sparallel}\rangle$ and $\sigma_\gamma^{\sparallel}$ are also available in the case of the extensible freely jointed chain model \cite{balabaev2009extension,fiasconaro2019analytical,buche2022freely}.
Eq.~\eqref{eq:gamma:longitudinal:s} is shown in Fig.~\ref{fig:gamma:longitudinal:s} as a function of the applied nondimensional force $\eta$ and varying numbers of links $N_b$.
The standard deviation of the longitudinal extension decreases steadily as the applied force increases or the number of links increases.
It is also important to consider the standard deviation $\sigma_\gamma^{\sparallel}$ scaled by the average $\langle\gamma_{\sparallel}\rangle$ to understand relative fluctuations \cite{mcquarrie}, shown by the dotted lines in Fig.~\ref{fig:gamma:longitudinal:s}.
Since the ensemble average longitudinal extension tends to zero as the force vanishes, but the standard deviation does not, relative fluctuations become asymptotically infinite.
This seems to indicate that for small forces, it could be difficult to measure the longitudinal extension of polymer chains with any certainty.

\begin{figure}[t]
\includegraphics{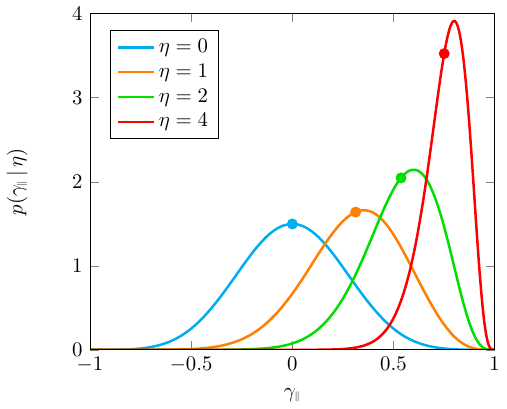}
\caption{\label{fig:gamma:longitudinal:p}%
Probability density $p$ of the longitudinal extension $\gamma_{\sparallel}$ under applied nondimensional force $\eta$ for freely jointed chains with 5 links.
The locations along the curves of the ensemble average longitudinal extension $\langle\gamma_{\sparallel}\rangle$ are also shown.
}
\end{figure}

In addition to moments such as the mean and variance, the underlying probability distribution can be examined.
The probability density of a longitudinal extension $\gamma_{\sparallel}$ given an applied force $\eta$ is the ensemble average of states with that particular longitudinal extension:
\begin{equation}\label{eq:gamma:longitudinal:p}
p(\gamma_{\sparallel}\,|\,\eta) = \left\langle\delta\left(\frac{1}{N_b}\sum_{i=1}^{N_b}\cos\theta_i - \gamma_{\sparallel}\right)\right\rangle.
\end{equation}
While this distribution cannot be calculated analytically, it is calculated numerically by randomly generating a large number of random chain configurations and binning the $z$ component of the extension \cite{conspire}, since $\gamma_{\sparallel}=\gamma_z$.
Eq.~\eqref{eq:gamma:longitudinal:p} is shown in Fig.~\ref{fig:gamma:longitudinal:p} as a function of the longitudinal extension $\gamma_{\sparallel}$ and varying nondimensional force $\eta$.
Initially broad and centered at $\gamma_{\sparallel}=0$, the distribution narrows and translates toward higher $\gamma_{\sparallel}$ as $\eta$ increases.
As evidenced by Fig.~\ref{fig:gamma:longitudinal:s}, the trends in Fig.~\ref{fig:gamma:longitudinal:p} show there are still considerable fluctuations in the longitudinal extension as the force increases.
These fluctuations are important to understand, since Eq.~\eqref{eq:gamma:longitudinal:m} is widely utilized in the literature without regard for uncertainty.
Also shown in Fig.~\ref{fig:gamma:longitudinal:p} is the location along the probability density of the ensemble average value given by Eq.~\eqref{eq:gamma:longitudinal:m}.

\subsection{Lateral extension}\label{sec:extension:lateral}

\begin{figure}[t]
\includegraphics{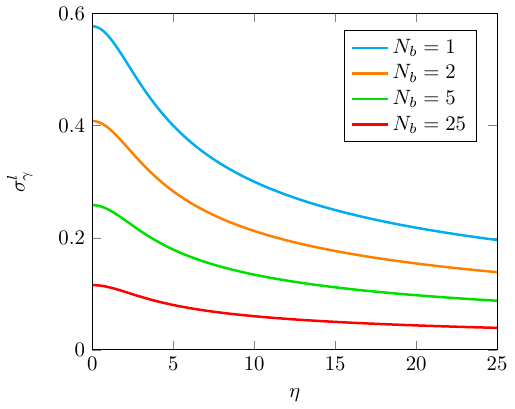}
\caption{\label{fig:gamma:lateral:s}%
Standard deviation $\sigma_\gamma^l$ of the lateral extension $\gamma_l$ as a function of the applied nondimensional force $\eta$ for freely jointed chains with numbers of links $N_b$.
Note that $\langle\gamma_l\rangle$, the ensemble average lateral extension, is exactly zero.
}
\end{figure}

Another component of the extension vector $\boldsymbol{\gamma}$ is the lateral extension $\gamma_l$, defined as the extension along any axis orthogonal to the nondimensional force vector $\boldsymbol{\eta}$.
With the force applied along the polar axis, it is
\begin{equation}\label{eq:gamma:lateral}
\gamma_l = \frac{1}{N_b}\sum_{i=1}^{N_b}\sin\theta_i\cos(\phi_i - \phi_0),
\end{equation}
where $\phi_0$ is azimuthal angle offset.
Since the polar axis is the $z$ axis, $\phi_0=0$ yields $\gamma_x$ and $\phi_0=\tfrac{\pi}{2}$ yields $\gamma_y$.
Further, the ensemble average of $\cos(\phi_i - \phi_0)$ is zero for all $\phi_i$ and any $\phi_0$, so the average lateral extension is zero,
\begin{equation}\label{eq:gamma:lateral:m}
\langle\gamma_l\rangle = 0.
\end{equation}
This is an intuitive result, because an applied force should not cause net bias to the extension of freely jointed chain on axes which are orthogonal to the force, by definition.
However, fluctuations around $\langle\gamma_l\rangle=0$ are non-trivial and a function of the applied force.
Since the averages are zero, mixed terms in the variance are zero, allowing
\begin{equation}
\langle\gamma_l^2\rangle = \frac{1}{N_b^2}\left\langle\sum_{i=1}^{N_b}\sin^2\!\theta_i\cos^2(\phi_i - \phi_0)\right\rangle.
\end{equation}
This can be exactly integrated to $\mathcal{L}(\eta)/\eta/N_b$, so the standard deviation of the lateral extension is therefore
\begin{equation}\label{eq:gamma:lateral:s}
\sigma_\gamma^l(\eta) = \sqrt{\frac{1}{N_b}\left(\frac{1}{\eta\tanh\eta} - \frac{1}{\eta^2}\right)}.
\end{equation}
Eq.~\eqref{eq:gamma:lateral:s} is shown in Fig.~\ref{fig:gamma:lateral:s} as a function of the applied nondimensional force $\eta$ and varying numbers of links $N_b$.
As the applied force increases, the standard deviation of the lateral extension decreases much more slowly than that of the longitudinal extension in Fig.~\ref{fig:gamma:longitudinal:s}.
Therefore, while a sufficiently large force would make the longitudinal position of the chain end somewhat certain, the chain end could still exhibit substantial lateral fluctuations.
In either case, the standard deviation has the same $N_b^{-1/2}$ dependence which decreases fluctuations as the number of links in the chain increases.
Finally, there is no relative standard deviation shown in Fig.~\ref{fig:gamma:lateral:s} because the ensemble average lateral extension is always zero, Eq.~\eqref{eq:gamma:lateral:m}.

\begin{figure}[t]
\includegraphics{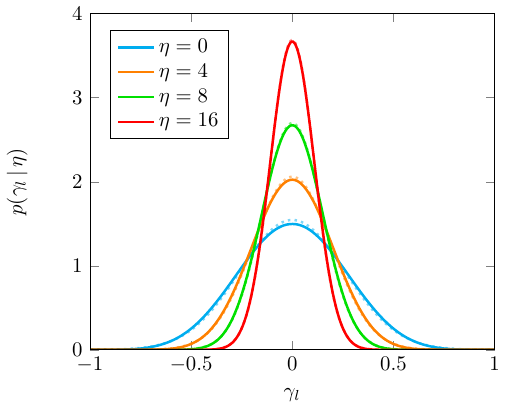}
\caption{\label{fig:gamma:lateral:p}%
Probability density $p$ of the lateral extension $\gamma_l$ under applied nondimensional force $\eta$ for freely jointed chains with 5 links.
Dotted lines show the Gaussian distributions associated with zero mean $\langle\gamma_l\rangle=0$ and standard deviation $\sigma_\gamma^l$.
}
\end{figure}

Similar to Eq.~\eqref{eq:gamma:longitudinal:p}, the probability density $p(\gamma_l\,|\,\eta)$ is the ensemble average of states with a lateral extension given by Eq.~\eqref{eq:gamma:lateral}, where $\phi_0=0$ is chosen for simplicity:
\begin{equation}\label{eq:gamma:lateral:p}
p(\gamma_l\,|\,\eta) = \left\langle\delta\left(\frac{1}{N_b}\sum_{i=1}^{N_b}\sin\theta_i\cos\phi_i - \gamma_l\right)\right\rangle.
\end{equation}
This distribution also cannot be calculated analytically, so it is calculated numerically in the same way as Eq.~\eqref{eq:gamma:longitudinal:p} except the $x$ component is binned \cite{conspire}.
Eq.~\eqref{eq:gamma:lateral:p} is shown in Fig.~\ref{fig:gamma:lateral:p} as a function of the lateral extension $\gamma_l$ and varying nondimensional force $\eta$.
The distribution of lateral extensions $\gamma_l$ both narrows and rises as $\eta$ increases, always remaining symmetric about $\gamma_l=0$.
This distribution narrows much more slowly with increasing force than the distribution of longitudinal extensions in Fig.~\ref{fig:gamma:longitudinal:p}, which was expected from comparing the standard deviations in Figs.~\ref{fig:gamma:longitudinal:s} and \ref{fig:gamma:lateral:s} in the discussion above.
Additionally shown in Fig.~\ref{fig:gamma:lateral:p} are the Gaussian approximations of the distribution using Eqs.~\eqref{eq:gamma:lateral:m} and \eqref{eq:gamma:lateral:s}.
Even with only 5 links in the chain, the distribution of lateral extensions appears to be nearly Gaussian, an approximation that improves as the applied force increases.

\subsection{Transverse extension}\label{sec:extension:transverse}

\begin{figure}[t]
\includegraphics{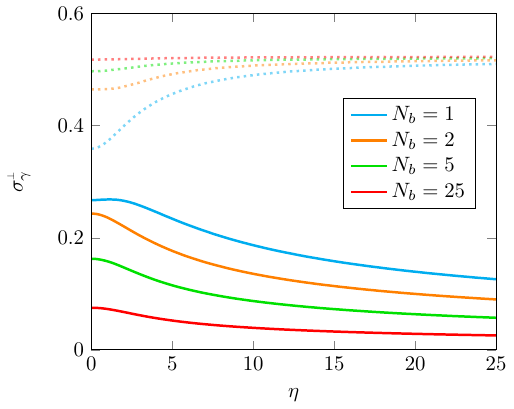}
\caption{\label{fig:gamma:transverse:s}%
Standard deviation $\sigma_\gamma^{\sperp}$ of the transverse extension $\gamma_{\sperp}$ as a function of the applied nondimensional force $\eta$ for freely jointed chains with numbers of links $N_b$.
Dotted lines show $\sigma_\gamma^{\sperp}$ relative to $\langle\gamma_{\sperp}\rangle$, the ensemble average transverse extension.
}
\end{figure}

The magnitude of the extension vector $\boldsymbol{\gamma}$ perpendicular to the nondimensional force vector $\boldsymbol{\eta}$ is defined as the transverse extension $\gamma_{\sperp}$, which is then written as
\begin{equation}\label{eq:gamma:transverse}
\gamma_{\sperp} = \sqrt{\gamma^2 - \gamma_{\sparallel}^2} = \sqrt{\gamma_x^2 + \gamma_y^2}.
\end{equation}
While the lateral extension $\gamma_l$ is the magnitude of the extension along any one axis orthogonal to the applied force, the transverse extension $\gamma_{\sperp}$ is the magnitude of the extension in the plane orthogonal to the applied force.
Further, while $\langle\gamma_l\rangle = 0$, here $\langle\gamma_{\sperp}\rangle \neq 0$.
Even further, while the ensemble average values and standard deviations have been obtained analytically up to this point, now they can no longer be obtained analytically.
Instead, the probability density distribution $p(\gamma_{\sperp}\,|\,\eta)$ in Eq.~\eqref{eq:gamma:transverse:p} is integrated numerically for both the ensemble average transverse extension
\begin{equation}\label{eq:gamma:transverse:m}
\langle\gamma_{\sperp}\rangle = \int_0^1 p(\gamma_{\sperp}\,|\,\eta)\,\gamma_{\sperp}\,d\gamma_{\sperp},
\end{equation}
and the standard deviation from the variance
\begin{equation}\label{eq:gamma:transverse:s}
(\sigma_\gamma^{\sperp})^2 = \int_0^1 p(\gamma_{\sperp}\,|\,\eta)\,\gamma_{\sperp}^2\,d\gamma_{\sperp} - \langle\gamma_{\sperp}\rangle^2.
\end{equation}
Unfortunately, unlike the cases that will follow, it seems that $p(\gamma_{\sperp}\,|\,\eta)$ cannot be formulated analytically either.
The distribution itself is still calculated using one Monte Carlo calculation for each curve, like Figs.~\ref{fig:gamma:longitudinal:p} and \ref{fig:gamma:lateral:p}, but now one calculation is needed for every point in the curves plotted for the standard deviation in Eq.~\eqref{eq:gamma:transverse:s}.
Therefore, the transverse extension appears to be the most computationally intensive component to examine.

Eq.~\eqref{eq:gamma:transverse:s} is shown in Fig.~\ref{fig:gamma:transverse:s} as a function of the applied nondimensional force $\eta$ and varying numbers of links $N_b$.
Ten million Monte Carlo samples are used to calculate each of the three hundred points along each curve \cite{conspire}.
As the applied force increases, the standard deviation of the transverse extension decreases slowly, similar to that of the lateral extension in Fig.~\ref{fig:gamma:longitudinal:s}.
Note that the relative standard deviation $\sigma_\gamma^{\sperp}/\langle\gamma_{\sperp}\rangle$ in Fig.~\ref{fig:gamma:transverse:s} approaches a constant as $\eta$ increases, the same constant for any $N_b$.
This is because for $\eta\gg 1$, the distribution of transverse extensions becomes a Rayleigh distribution \cite{strutt1919problem}, and Rayleigh distributions have constant relative standard deviations, so relative fluctuations would be ubiquitous.

\begin{figure}[t]
\includegraphics{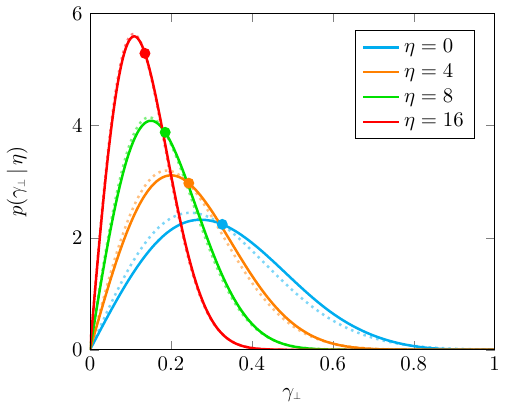}
\caption{\label{fig:gamma:transverse:p}%
Probability density $p$ of the transverse extension $\gamma_{\sperp}$ under applied nondimensional force $\eta$ for freely jointed chains with 5 links.
Dotted lines show Rayleigh distribution approximations, and the locations of $\langle\gamma_{\sperp}\rangle$ are also shown.
}
\end{figure}

The probability density $p(\gamma_{\sperp}\,|\,\eta)$ is the ensemble average of states with the transverse extension $\gamma_{\sperp}$
\begin{equation}\label{eq:gamma:transverse:p}
p(\gamma_{\sperp}\,|\,\eta) = \left\langle\delta\left(\sqrt{|\boldsymbol{\gamma}|^2 - \gamma_{\sparallel}^2} - \gamma_{\sperp}\right)\right\rangle.
\end{equation}
where the longitudinal extension $\gamma_{\sparallel}$ is given configurationally by Eq.~\eqref{eq:gamma:longitudinal}.
This distribution is calculated numerically the same way as Eq.~\eqref{eq:gamma:lateral:p} except the norm of the extension vector in the $xy$ plane is binned \cite{conspire}.
Eq.~\eqref{eq:gamma:transverse:p} is shown in Fig.~\ref{fig:gamma:transverse:p} as a function of the transverse extension $\gamma_{\sperp}$ and varying nondimensional force $\eta$.
The distribution of transverse extensions $\gamma_{\sperp}$ gradually narrows and moves towards $\gamma_{\sperp} = 0$ as $\eta$ increases.
Also shown in Fig.~\ref{fig:gamma:transverse:p} are the Rayleigh approximations of the distribution via
\begin{equation}\label{eq:gamma:transverse:p:lim}
p(\gamma_{\sperp}\,|\,\eta) = \frac{\gamma_{\sperp}}{\sigma^2}\exp\left(-\frac{\gamma_{\sperp}^2}{2\sigma^2}\right)\text{ for }\eta\gg 1,
\end{equation}
where $\sigma^2 = 2 (\sigma_\gamma^{\sperp})^2 / (4 - \pi)$.
This approximation is quite accurate and improves as the applied force increases.

\subsection{Radial extension}\label{sec:extension:radial}

\begin{figure}[t]
\includegraphics{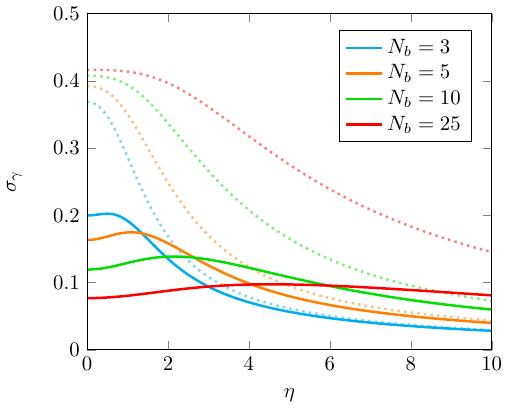}
\caption{\label{fig:gamma:radial:s}%
Standard deviation $\sigma_\gamma$ of the radial extension $\gamma$ as a function of the nondimensional force $\eta$ for freely jointed chains with numbers of links $N_b$.
Dotted lines show $\sigma_\gamma$ relative to $\langle\gamma\rangle$, the ensemble average radial extension.
}
\end{figure}

The absolute value of the extension vector $\boldsymbol{\gamma}$ is defined as the radial extension, given simply by
\begin{equation}\label{eq:gamma:radial}
\gamma = |\boldsymbol{\gamma}| = \sqrt{\gamma_{\sparallel}^2 + \gamma_{\sperp}^2},
\end{equation}
which is the nondimensional end-to-end length per link.
Similar to the transverse extension, neither the ensemble average radial extension
\begin{equation}\label{eq:gamma:radial:m}
\langle\gamma\rangle = \int_0^1 p(\gamma\,|\,\eta)\,\gamma\,d\gamma,
\end{equation}
nor the standard deviation, from the variance
\begin{equation}\label{eq:gamma:radial:s}
(\sigma_\gamma)^2 = \int_0^1 p(\gamma\,|\,\eta)\,\gamma^2\,d\gamma - \langle\gamma\rangle^2,
\end{equation}
can be obtained analytically.
Instead, they are integrated numerically using the relation for $p(\gamma\,|\,\eta)$ in Eq.~\eqref{eq:gamma:radial:p}.
Critically, notice that $\langle\gamma\rangle\neq\mathcal{L}(\eta)$ because the Langevin function actually governs the only longitudinal component of the extension, as in Eq.~\eqref{eq:gamma:longitudinal:m}.
This would be an easy mistake to make since nearly all sources in the literature do not delineate scalar extension measures.
Eq.~\eqref{eq:gamma:radial:s} is shown in Fig.~\ref{fig:gamma:radial:s} as a function of the applied nondimensional force $\eta$ and varying numbers of links $N_b$.
As the applied force $\eta$ increases, the standard deviation of the radial extension $\sigma_\gamma$ slightly increases, before decreasing as the force continues to increase afterwards.
Both the rate of increase and following decrease of $\sigma_\gamma$ gradually diminishes as the number of links $N_b$ grows, causing the crossover seen in Fig.~\ref{fig:gamma:radial:s}.
The relative standard deviation $\sigma_\gamma/\langle\gamma\rangle$ is also shown, which decreases over all $\eta$.

\begin{figure}[t]
\includegraphics{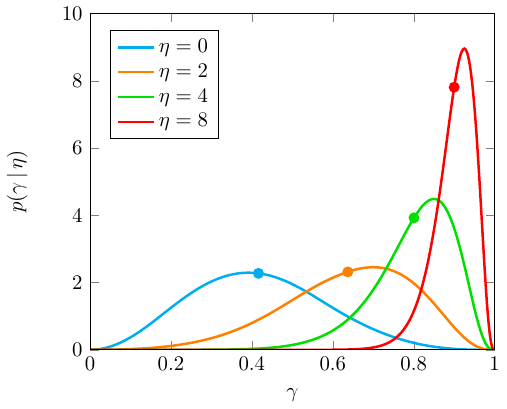}
\caption{\label{fig:gamma:radial:p}%
Probability density $p$ of the radial extension $\gamma$ under applied nondimensional force $\eta$ for freely jointed chains with 5 links.
The locations along the curves of the ensemble average radial extension $\langle\gamma\rangle$ are also shown.
}
\end{figure}

The isotensional partition function in Eq.~\eqref{eq:z} can be formulated in terms of the isometric partition function $Q$ through a Laplace transform \cite{buche2020statistical}, which would be
\begin{equation}\label{eq:z:q}
Z(\boldsymbol{\eta}) = \iiint Q(\boldsymbol{\gamma})\,e^{N_b\boldsymbol{\eta}\cdot\boldsymbol{\gamma}}\,d^3\boldsymbol{\gamma}.
\end{equation}
Both partition functions are spherically symmetric, so scalar arguments are used and the integrals over both the polar and azimuthal angles are computed, leaving
\begin{equation}\label{eq:z:gamma}
Z(\eta) = 4\pi\int Q(\gamma)\sinhc(N_b\eta\gamma)\,\gamma^2\,d\gamma,
\end{equation}
where $\sinhc$ is the hyperbolic sinc function.
Since the radial distribution function $g(\gamma)$ is related \cite{buche2020statistical} to the isometric partition function through $g(\gamma)\propto 4\pi\gamma^2 Q(\gamma)$, and since $p(\gamma\,|\,\eta) = \langle\delta(|\boldsymbol{\gamma}| - \gamma)\rangle$, the probability density of the radial extension $\gamma$ under an applied force $\eta$ is
\begin{equation}\label{eq:gamma:radial:p}
p(\gamma\,|\,\eta) \propto g(\gamma)\sinhc(N_b\eta\gamma).
\end{equation}
The radial distribution function \cite{treloar1949physics,wang1952statistical,buche2023modeling} is given by
\begin{equation}\label{eq:g}
g(\gamma) = \frac{\gamma}{2}\frac{N_b^{N_b}}{(N_b - 2)!}\sum_{s=0}^{s_\mathrm{max}}(-1)^s\tbinom{N_b}{s}\left(m - \tfrac{s}{N_b}\right)^{N_b - 2},
\end{equation}
where $m=(1-\gamma)/2$ and $s_\mathrm{max}=\lfloor mN_b\rfloor$.
Eq.~\eqref{eq:gamma:radial:p} is shown in Fig.~\ref{fig:gamma:radial:p} as a function of the radial extension $\gamma$ and varying nondimensional force $\eta$.
The trends in radial extensions in Fig.~\ref{fig:gamma:radial:p} are similar to those in Fig.~\ref{fig:gamma:longitudinal:p} for the longitudinal extension, except the shape of $p(\gamma\,|\,\eta)$ is quite different and is on $\gamma\in[0,1]$, and it is also less sensitive to increasing force.
In either case, the limiting distribution for $\eta\gg 1$ case is not readily apparent.

\subsection{Extension angle}\label{sec:extension:angular}

\begin{figure}[t]
\includegraphics{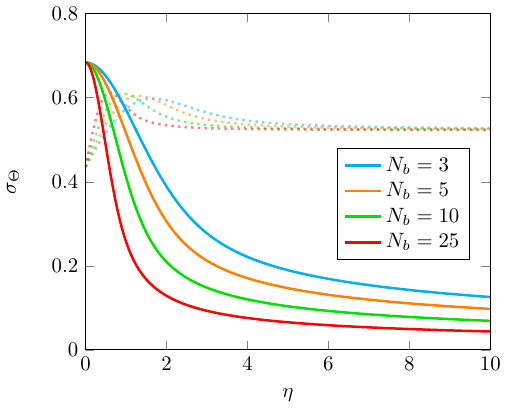}
\caption{\label{fig:gamma:angular:s}%
Standard deviation $\sigma_\Theta$ of extension angle $\Theta$ as a function of the nondimensional force $\eta$ for freely jointed chains with numbers of links $N_b$.
Dotted lines show $\sigma_\Theta$ relative to $\langle\Theta\rangle$, the ensemble average extension angle.
}
\end{figure}

The angle $\Theta$ between the extension vector $\boldsymbol{\gamma}$ and the nondimensional force vector $\boldsymbol{\eta}$ is defined through
\begin{equation}\label{eq:gamma:angular-0}
\boldsymbol{\eta}\cdot\boldsymbol{\gamma}=\eta\,\gamma\cos\Theta,
\end{equation}
where the radial extension is $\gamma=|\boldsymbol{\gamma}|$.
The extension angle can also be related to the longitudinal extension by $\gamma_{\sparallel}=\gamma\cos\Theta$ or the transverse extension by $\gamma_{\sperp}=\gamma\sin\Theta$, which permits any of the following relations to hold:
\begin{equation}\label{eq:gamma:angular}
\Theta = \sin^{-1}\left(\frac{\gamma_{\sperp}}{\gamma}\right) = \cos^{-1}\left(\frac{\gamma_{\sparallel}}{\gamma}\right) = \tan^{-1}\left(\frac{\gamma_{\sperp}}{\gamma_{\sparallel}}\right).
\end{equation}
The ensemble average extension angle is given by
\begin{equation}\label{eq:gamma:angular:m}
\langle\Theta\rangle = \int_0^\pi p(\Theta\,|\,\eta)\,\Theta\,d\Theta,
\end{equation}
and the standard deviation $\sigma_\Theta$, from the variance
\begin{equation}\label{eq:gamma:angular:s}
(\sigma_\Theta)^2 = \int_0^\pi p(\Theta\,|\,\eta)\,\Theta^2\,d\Theta - \langle\Theta\rangle^2,
\end{equation}
but neither of these relations can be obtained analytically, so they are instead integrated numerically using the relation for $p(\Theta\,|\,\eta)$ in Eq.~\eqref{eq:gamma:angular:p}.
Eq.~\eqref{eq:gamma:angular:s} is shown in Fig.~\ref{fig:gamma:angular:s} as a function of the applied nondimensional force $\eta$ and varying numbers of links $N_b$.
The standard deviation drops sharply as the applied force increases before experiencing a more gradual decrease.
Similar to the transverse extension in Fig.~\ref{fig:gamma:transverse:s}, the relative standard deviation becomes constant as the force becomes large.
This is because for $\eta\gg 1$, the distribution of extension angles in Eq.~\eqref{eq:gamma:angular:p} also becomes Rayleigh, i.e., Eq.~\eqref{eq:link:theta:p:lim}.

\begin{figure}[t]
\includegraphics{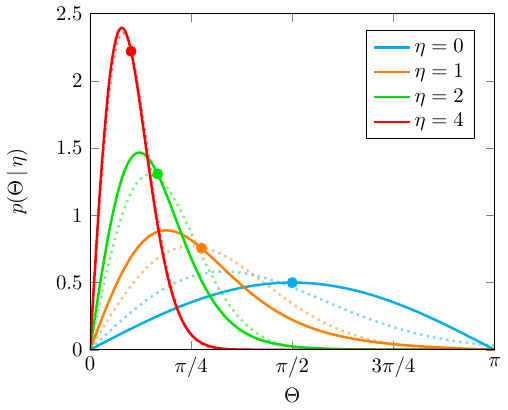}
\caption{\label{fig:gamma:angular:p}%
Probability density $p$ of the extension angle $\Theta$ under applied nondimensional force $\eta$ for freely jointed chains with 5 links.
Dotted lines show Rayleigh distribution approximations, and the locations of $\langle\Theta\rangle$ are also shown.
}
\end{figure}

In order to get from Eq.~\eqref{eq:z:q} to Eq.~\eqref{eq:z:gamma}, both the integrals over the azimuthal angle as well as the polar angle, which is the extension angle $\Theta$ via Eq.~\eqref{eq:gamma:angular-0}, were analytically computed.
Here, only the azimuthal angle integral is computed, which simply amounts to $2\pi$, leaving
\begin{equation}
Z(\eta) = 2\pi\iint Q(\gamma)\,e^{N_b\eta\gamma\cos\Theta}\,\gamma^2\,d\gamma\,\sin\Theta\,d\Theta.
\end{equation}
Again, since the radial distribution function $g(\gamma)$ given by Eq.~\eqref{eq:g} is related \cite{buche2020statistical} to the isometric partition function through $g(\gamma)\propto 4\pi\gamma^2 Q(\gamma)$, the probability density of the extension angle $\Theta$ under an applied force $\eta$ is
\begin{equation}\label{eq:gamma:angular:p}
p(\Theta\,|\,\eta) \propto \int_0^1 e^{N_b\eta\gamma\cos\Theta}\sin\Theta\,g(\gamma)\,d\gamma.
\end{equation}
In the case of the radial extension, the probability density in Eq.~\eqref{eq:gamma:radial:p} needs to be normalized numerically before being used, which is a single numerical quadrature computation.
Here, in the case of the extension angle, the probability density in Eq.~\eqref{eq:gamma:angular:p} still needs to be normalized, but now every evaluation of the probability density requires numerical quadrature since the remaining integral over the radial extension $\gamma$ cannot be evaluated analytically.
However, these are still more efficient and reliable than Monte Carlo calculations.
Eq.~\eqref{eq:gamma:angular:p} is shown in Fig.~\ref{fig:gamma:angular:p} as a function of the radial extension $\gamma$ and varying nondimensional force $\eta$.
The zero force distribution is $p(\Theta\,|\,0) = \sin\Theta/2$, an equal distribution on the sphere.
As the applied force increases, the distribution narrows and translates towards the limit at $\Theta=0$, faster than but consistent with the shrinking transverse extension in Fig.~\ref{fig:gamma:transverse:p}.
Also shown in Fig.~\ref{fig:gamma:angular:p} are the approximations verifying that Eq.~\eqref{eq:gamma:angular:p} becomes Rayleigh for $\eta\gg 1$.

\section{Chain link angles}\label{sec:angle}

\begin{figure}[t]
\includegraphics{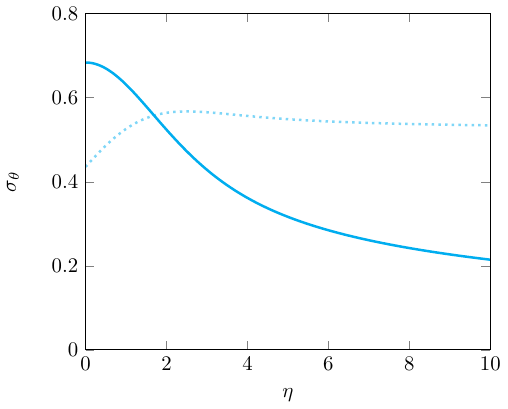}
\caption{\label{fig:link:theta:s}%
Standard deviation $\sigma_\theta$ of any link angle $\theta$ as a function of nondimensional force $\eta$ for freely jointed chains with any number of links.
Dotted line shows $\sigma_\theta$ relative to $\langle\theta\rangle$, the ensemble average of any link angle.
}
\end{figure}

The angle $\theta_i$ of each link $i=1,\ldots,N_b$ in the chain is defined as the angle between the link and the polar axis along which the nondimensional force $\eta$ is applied.
Note that the statistics of any given link angle $\theta$ are independent of the number of links in the chain $N_b$, and would exactly match those for the extension angle $\Theta$ if $N_b=1$.
Similarly, note that the statistics of $\cos\theta$ would exactly match those for the longitudinal extension $\gamma_{\sparallel}$ if $N_b=1$.
The ensemble average of any one link angle is given by
\begin{equation}\label{eq:link:theta:m}
\langle\theta\rangle = \int_0^\pi p(\theta\,|\,\eta)\,\theta\,d\theta,
\end{equation}
and the standard deviation $\sigma_\theta$, from the variance
\begin{equation}\label{eq:link:theta:s}
(\sigma_\theta)^2 = \int_0^\pi p(\theta\,|\,\eta)\,\theta^2\,d\theta - \langle\theta\rangle^2,
\end{equation}
which are integrated numerically using Eq.~\eqref{eq:link:theta:p}, the analytic relation for $p(\theta\,|\,\eta)$.
Eq.~\eqref{eq:link:theta:s} is shown in Fig.~\ref{fig:link:theta:s} as a function of the applied nondimensional force $\eta$.
As the applied force increases, the standard deviation briefly increases before gradually decreasing afterwards, similar to the extension angle in Fig.~\ref{fig:gamma:angular:s}.
The relative standard deviation $\sigma_\theta/\langle\theta\rangle$ also approaches the same constant $\sqrt{4/\pi - 1}$ for $\eta\gg 1$ since the distribution of link angles also becomes Rayleigh in that limit.

\begin{figure}[t]
\includegraphics{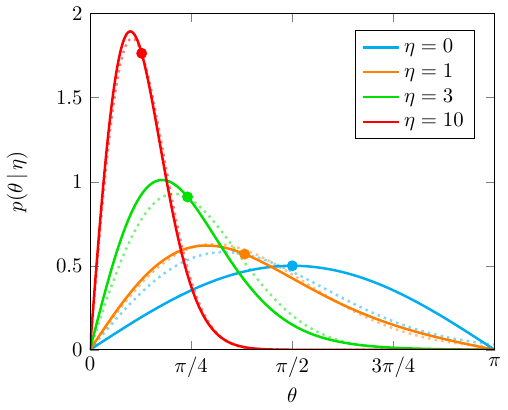}
\caption{\label{fig:link:theta:p}%
Probability density $p$ of any link angle $\theta$ from applied nondimensional force $\eta$ for freely jointed chains of any number of links.
Dotted lines show Rayleigh distribution approximations, and the locations of $\langle\theta\rangle$ are also shown.
}
\end{figure}

For $N_b=1$, the radial distribution function in Eq.~\eqref{eq:g} becomes $g(\gamma)=\delta(\gamma-1)$, so Eq.~\eqref{eq:gamma:angular:p} can be exactly integrated and normalized, leaving the probability density of any given link angle $\theta$ under an applied force $\eta$ as
\begin{equation}\label{eq:link:theta:p}
p(\theta\,|\,\eta) = \frac{e^{\eta\cos\theta}\sin\theta}{2\sinhc\eta}.
\end{equation}
Eq.~\eqref{eq:link:theta:p} is shown in Fig.~\ref{fig:link:theta:p} as a function of the radial extension $\gamma$ and varying nondimensional force $\eta$.
The behavior is quite similar to the extension angle in Fig.~\ref{fig:gamma:angular:p}, though, the width of the distribution of a link angle is generally less sensitive to the increasing force.
This is expected since distributions that depend on the number of links in the chain, such as the extension angle or any measure of extension, tend to narrow as the number of links in the chain increases \cite{buche2020statistical}.
The Rayleigh approximations of the distribution of link angles $p(\theta\,|\,\eta)$ in Eq.~\eqref{eq:link:theta:p} are also shown in Fig.~\ref{fig:link:theta:p}, which only start to become accurate in the expected limit of large $\eta$,
\begin{equation}\label{eq:link:theta:p:lim}
p(\theta\,|\,\eta) \sim \frac{\theta}{\sigma^2}\exp\left(-\frac{\theta^2}{2\sigma^2}\right)\text{ for }\eta\gg 1,
\end{equation}
where $\sigma^2 = 2 (\sigma_\theta)^2 / (4 - \pi)$.
The mean is $\sigma\sqrt{\pi/2}$, which is why the relative standard deviation $\sigma_\theta/\langle\theta\rangle=\sqrt{4/\pi - 1}$ is constant in this regime.
As expected $p(\theta\,|\,\eta)\to\delta(\theta)$ as $\eta\to\infty$, i.e., the links align as the chain is fully extended.

It is vital to emphasize the differences between the statistics of the link angles here in Sec.~\ref{sec:angle} and the statistics of the extension related quantities in Sec.~\ref{sec:extension}.
First, while the extension related quantities become more narrowly distributed as the number of links in the chain increases, the statistics of any given link remain the same.
Second, these would be the statistics of any one link in the chain, not statistics over all links in the chain (for example, the minimum link angle in the chain).
Finally, it would be inaccurate and naive to assume any one configuration for the links in the chain anytime before the applied force becomes considerably large, at which point the configuration of links would become trivial as they all approximately align.

\section{Conclusion}\label{sec:conclusion}

Fluctuations in the thermodynamic quantities of interest for the freely jointed chain model in the isotensional ensemble have been examined in detail.
These quantities included several relevant components of the chain extension, and the angle of a given link in the chain.
Absolute and relative standard deviations quantified the expected fluctuations of these quantities around their ensemble average values.
Probability density distributions further highlighted fluctuations in all quantities and their general statistics under increasing force.
While the statistical distribution of each component of the extension gradually narrows as the applied force or number of links in the chain increases, the statistics of any given link only narrows with an increasing force.
In the future, similarly diligent studies should be completed involving extensible links \cite{buche2022freely}, other single-chain models \cite{fiasconaro2023elastic}, more thermodynamic ensembles \cite{manca2012elasticity}, and compounding effects from imperfect stretching devices \cite{buche2023modeling}.

\begin{acknowledgments}
Sandia National Laboratories is a multi-mission laboratory managed and operated by National Technology and Engineering Solutions of Sandia, LLC., a wholly owned subsidiary of Honeywell International, Inc., for the U.S. Department of Energy's National Nuclear Security Administration under Contract No. DE-NA0003525. Any subjective views or opinions expressed in the paper do not necessarily represent the views of the U.S. Department of Energy or the U.S. Government. The U.S. Government retains and the publisher, by accepting the article for publication, acknowledges that the U.S. Government retains a nonexclusive, paid-up, irrevocable, world-wide license to publish or reproduce the published form of this manuscript, or allow others to do so, for U.S. Government purposes.
\end{acknowledgments}

\bibliography{main}

\end{document}